# A comprehensive review of sensor technologies, instrumentation, and signal processing solutions for low-power Internet of Things systems with mini-computing devices

Alexandros Gazis[1,*], Ioannis Papadongonas[2], Athanasios Andriopoulos[3], Constantinos Zioudas[4], Theodoros Vavouras[5,6]

## Abstract

This article provides a comprehensive overview of sensors commonly used in low-cost, low-power systems, focusing on key concepts such as Internet of Things (IoT), Big Data, and smart sensor technologies. It outlines the evolving roles of sensors, emphasizing their characteristics, technological advancements, and the transition toward "smart sensors" with integrated processing capabilities. The article also explores the growing importance of mini-computing devices in educational environments. These devices provide cost-effective and energy-efficient solutions for system monitoring, prototype validation, and real-world application development. By interfacing with wireless sensor networks and IoT systems, mini-computers enable students and researchers to design, test, and deploy sensor-based systems with minimal resource requirements. Furthermore, this article examines the most widely used sensors, detailing their properties and modes of operation to help readers understand how sensor systems function. The aim of this study is to provide an overview of the most suitable sensors for various applications by explaining their uses and operations in simple terms. This clarity will assist researchers in selecting the appropriate sensors for educational and research purposes or understanding why specific sensors were chosen, along with their capabilities and possible limitations. Ultimately, this research seeks to equip future engineers with the knowledge and tools needed to integrate cutting-edge sensor networks, IoT, and Big Data technologies into scalable, real-world solutions.

## 1. Introduction

In recent years, Internet of Things (IoT) and Big Data have been extensively incorporated into the education of young scientists [1]. The first term is derived from the ever-increasing number of devices that are connected to a large interconnected network of computing devices [2]. The second term is derived from the data that are increasing in our everyday lives and applications, making them so "big" that there needs to be a specific term to categorize them [3]. The term "big" is used to denote terms of volume, velocity, and variety (the three Vs) [4], which is closely associated to their processing, which is performed using deep learning techniques or Artificial Intelligence (AI).

One of the most basic ways of understanding and finding all this information is the so-called intelligent sensors (smart sensors) [5]. As such, the future of these technology domains is interconnected, with the main issue being that the processing power of testing new and existing methods on these areas is becoming exponential. For example, the new trend of chatbot machines using AI such as generative adversarial networks (GAN), which is one of the hottest trends and widely used applications of AI, has been heavily criticized for both training the data used and the time and electricity required to answer even the simplest of questions [6]. As such, there needs to be a computing device that will be able to be used in schools and research facilities as a rapid prototype solution or as a training tool for the students to be educated and learn about these new technologies [7–9]. These devices, if focused in all of the areas mentioned above, must be both low power and low cost so that either students/young scientists [10] or hobby enthusiasts, aiming for cost-effectiveness [11], can and will be able to use them at school or research facilities. They will also be able to easily buy/replace and maintain them at the lowest cost possible [12–14]. To find this threshold between pricing and computing capabilities, several solutions are being discovered every year [15, 16]. Some examples include new mini-computers that can connect to external computing devices and support even big and resource-intensive applications such as computer vision [17, 18]. The main devices studied and presented here are not microcontrollers, which receive a signal and base some code/process to provide a single output, but are smarter devices that also incorporate some sort of feedback and memory for the programmers/electrical engineers [19].

[1] Department of Electrical and Computer Engineering, School of Engineering, Democritus University of Thrace, 67100 Xanthi, Greece.
[2] Department of Electrical Engineering and Information Technology, School of Computation, Information and Technology, 80333, Munich, Germany.
[3] Department of Business Administration, School of Administrative, Economics and Social Sciences, University of West Attica, 12243 Athens, Greece.
[4] Department of Social Sciences, School of Social Sciences, Hellenic Open University, 26331 Patra, Greece.
[5] Department of Humanities, School of Humanities, Hellenic Open University, 26335 Patra, Greece.
[6] Department of Theoretical and Applied Linguistics, School of Italian Language and Literature, Aristotle University of Thessaloniki, 54124 Thessaloniki, Greece.
*email: agazis@ee.duth.gr

As such, regardless of the needs or an application, the first step for researchers aiming for cost-effectiveness to get accustomed to these technologies is to have small and low-cost devices that will be able to support these applications (not necessarily at a scale) [20–22]. This article showcases the existing mini-computing devices and how to program small- to medium-sized applications. It suggests low-cost and low-power solutions mainly focusing on educational purpose applications. Specifically, this article first describes a sensor and briefly explains its properties and characteristics. Then, it highlights its applications and categories. It expands on the existing solutions where mini-computing devices can be used to connect, use, and host these sensors. It also provides more detailed comparison tables of computing devices that incorporate all the necessary skills for development. This article doesn't discuss about the pricing of sensors or mini-computing devices, as it is highly dependent on the area of purchase. However, it notes that the range of these devices and their respective capabilities are similar in terms of processing power.

## 2. Problem formulation

### 2.1. Defining sensory devices

A sensor is a device used to measure or detect a physical quantity and produce a measurable output. The first use of sensors started when they developed in living beings as their organs. Specifically, the human eyes and ears are typical examples through which we came to know what sensing is. Here, the former detects the spectrum of electromagnetic radiation and the latter detects sound, i.e., pressure, waves. Over time, instruments for solving everyday practical problems, such as measuring length, weight, and volume, were created [23, 24]. Later on, various observations and practical reasons in our everyday lives created the need for more accurate measurement than just assessing these physical quantities.

### 2.2. Defining sensory device generations and advancements

Since the beginning of sensor development, the term "smart sensor" has been used for various devices. This term refers to devices that have a fully or partially integrated information-processing unit. It is worth indicating that this embeddedness is necessary—either in the form of a data-processing system or in the form of memory feedback, an automatic calibration or compensation process, or even noise cancellation—for a sensor to be considered "smart" or "intelligent" [25–27].

First-generation "intelligent" sensors are usually connected to electronic signal processing and amplification circuitry [26, 28]. Second-generation sensors are located in their installation site and are remotely connected to a section of analog electronic circuitry to control their operations [29, 30]. Third-generation sensors contain a powerful sensor component usually connected with a signal determination module and are composed of integrated circuits and/or passive components existing in the same implementation part (module). The analog-to-digital (A/D) signal conversion in the converter and the microprocessor is an external element of the sensor composition and structure [31–33]. Fourth-generation smart sensors are a product of regulation circuits combined with an identical monolithic or hybrid integrated circuit. More specifically, in this phase, the transducer and digital processing circuits communicate with discrete elements and are, as in the previous generation, external elements of the sensor composition and signal conditioning circuits. The generated output is bidirectionally interfaced to a microprocessor that provides automatic control of operations [34–36].

Finally, in fifth-generation sensors, the converter of the A/D signal is located in a similar monolithic or hybrid integrated circuit in which the signal conditioner is placed. It is worth mentioning that depending on the design, a number of such sensors can have a digital signal as an output with the possibility of simultaneous and continuous communication with the microcontroller and the corresponding modern computer system. To achieve this function during their communication, a host system is used via a communication bus or a wired network [37, 38]. The main advantages of this generation are the existence of multiple signals from different sensors; automatic detection of the level of properties such as temperature, humidity, and other environmental factors that can disturb a measurement; the automatic correction of the main errors that occur during the operation of the predetermined life span of the components; and, in general, the integration of large-scale integrated systems (VLSI) [39].

### 2.3. Defining sensory device properties

For choosing an appropriate instrument for a specific application, it is important to know the characteristics of a sensor device. This is reflected by its performance and behavior during measurements. Some of the most important aspects to consider for technical instruments are as follows:

1. *Accuracy*: measuring how close is the measurement of the sensory device to the actual value of the property that is being measured. As such, high accuracy is translated to minimal error and reliable and accurate results for varying conditions [40].

2. *Tolerance*: refers to the acceptable range of deviation from a specified value of the values and conditions the sensor

can withstand without failing or producing incorrect readings [41].

3. *Linearity*: refers to the degree to which the sensor's output is directly proportional to the input across its entire range. As such, high linearity provides consistent and predictable measurements, whereas it may introduce errors and noise to the final data interpretation [42, 43].

4. *Distinctness*: refers to a sensor's ability to differentiate the values between small changes in the measured parameter. As such, sensors with high distinctness can detect fine variations in the input signal.

5. *Repeatability*: refers to the ability of a sensor to provide the same measurement results under the same conditions over multiple trials, thus ensuring reliability and consistent performance [44].

6. *Sensitivity*: refers to the sensor's ability to detect small changes in an input parameter. As such, a sensor with high sensitivity provides minimal variations, thus ensuring long-term monitoring of crucial environmental and operational changes and conditions [45].

## 2.4. Most known and widely used types of sensors

This section offers a comprehensive overview of various sensor types commonly used in measurement and control applications. We will explore sensors designed to measure temperature, optics, electrical resistivity, thermistors, pressure, rubber, capacitance, level, humidity, speed, distance, and force/weight. These sensors have vital applications across numerous industries, including manufacturing, automation, environmental monitoring, and scientific research. In the last subsection, we present two tables specifying various properties of the sensor types presented. **Table 1** focuses on the most known and used low-cost and low-power sensor devices, whereas **Table 2** expands this analysis and showcases a comparison of the most known and used low-cost and low-power computing devices in the industry.

**Table 1** • A detailed analysis of the most known and used low-cost and low-power sensor devices

| Sensor type | Ref. num. | State-of-the-art | Technology used | Computing devices used | Computing and signal processing | Power | Challenges and open issues | Interfacing and networking capacities |
|---|---|---|---|---|---|---|---|---|
| Temperature sensors | [23, 46, 47] | High precision, fast response | RTD, thermocouples, thermistors | Raspberry Pi, Arduino, ESP32 | Low to moderate | Low to moderate | Environmental drift, accuracy loss over time | I2C, SPI, Analog |
| Contact thermometers | [46, 47] | Direct contact, stable readings | Resistive materials, thermocouples | Arduino, Raspberry Pi | Low | Low | Limited range, mechanical wear | Analog, I2C |
| Remote thermometers | [48, 49] | Infrared or non-contact based solutions | IR sensors, optical detectors | ESP8266, Raspberry Pi | Moderate | Moderate | Calibration challenges, interference | Wireless, I2C |
| Optic sensors | [50, 51] | High-speed detention and precision | Fiber-optic, photodiodes | Jetson Nano, Raspberry Pi | High | Moderate | External light interference, complexity | Analog, Digital, USB |
| Electrical resistivity sensors | [52, 53] | Highly sensitive, low noise | Conductive materials, MEMS | STM32, Raspberry Pi | Moderate to high | Moderate | Temperature dependency, drift | Analog, I2C |
| Thermistor sensors | [54, 55] | Wide range, nonlinear response | Semiconductor oxides | Arduino, ESP32 | Low | Low | Nonlinear output, aging effects | Analog, I2C |
| Pressure sensors | [56, 57] | High sensitivity, MEMS integration | Piezoelectric, capacitive, resistive | Raspberry Pi, Industrial controllers | Moderate | Low to moderate | Signal drift, temperature dependency | I2C, SPI, Analog |
| Humidity sensors | [58, 59] | Capacitive or resistive sensing | Capacitive polymer, resistive films | Arduino, ESP8266 | Low | Low | Accuracy affected by contamination, response time | I2C, Analog |
| Speed sensors | [60, 61] | Hall effect or optical-based | Magnetic, optical encoders | STM32, Raspberry Pi | Moderate | Low | Noise interference, mechanical limitations | Digital, PWM |
| Distance sensors | [62, 63] | Ultrasonic or LIDAR | Sona, infrared, laser | Arduino, Raspberry Pi | High | Moderate | Environmental interference, | I2C, Serial, Analog |

| | | | | | | | | accuracy vs range trade-off | |
| Force-weight sensors | [64, 65] | Strain gauge based | Wheatstone Bridge, MEMS | Arduino, ESP32 | Moderate to high | Moderate | Drift over time, temperature compensation | Analog, Digital |

**Table 2** • A comparison analysis of the most known and used low-cost and low-power computing devices in the industry

| Device | CPU model | CPU technology | RAM | Speed | Power | Operating systems | Recommended programming languages | GFLOPS |
|---|---|---|---|---|---|---|---|---|
| Raspberry Pi 4 Model B | Quad-core 1.5GHz Arm Cortex-A72 | ARMv8-A | 1–8 GB LPDDR4 | 1.5 GHz | 5V 3A | Raspberry Pi OS, Ubuntu | Python, C, C++, Java, Scratch | 48.0 |
| Raspberry Pi 3 Model B | Quad-core 1.2GHz Broadcom BCM2837 | ARMv8-A (32-bit) | 1GB LPDDR2 | 1.2 GHz | 5V 2.5A | Raspberry Pi OS, Ubuntu | Python, C, C++, Java, Scratch | 38.4 |
| Onion Omega2+ | 580 MHz MIPS | MIPS 24KEc | 128MB DDR2 | 580 MHz | 3.3V 0.18A | OpenWrt, Debian | Python, JavaScript, C++ | 2.32 |
| ASUS Tinker Board S | Quad-core 1.8 GHz RK3288-CG.W | ARM Cortex-A17 | 2GB LPDDR3 | 1.8 GHz | 5V 1.6A | TinkerOS, Armbian | Python, C, C++, Java | 57.6 |
| Nvidia Jetson Nano | Quad-core ARM Cortex-A57 | ARMv8-A | 4GB LPDDR4 | 921 MHz | 5V 2A | Ubuntu-based JetPack OS: Linux4Tegra, Jetson Linux, Ambian | Python, C, C++, CUDA | 29.48 |

### 2.4.1. Sensors for measuring temperature

Temperature is defined as the physical quantity that determines the equilibrium of a system in terms of its thermal characteristics. The basic discovery for measuring this quantity was the thermometer, which nowadays consists mainly of electronic components and is divided into two categories:

1. *Contact thermometers*: these can produce the desired reading by coming into contact with the system whose temperature is being measured, i.e., by measuring their temperature. In this category, the accuracy of the measurement depends to a large extent on the extent to which thermal equilibrium has been established between the thermometer and the system [46].

2. *Remote thermometers*: these provide the desired indication of the thermal radiation of the system and indirectly calculate the temperature because physical contact between the thermometer and the system whose temperature is to be measured is not considered necessary [47].

The type of sensor to be used to obtain the required measurement depends on several factors, such as the range of variation in the temperature to be measured, the required accuracy, and the fidelity of the environment in which the sensor is placed. Mechanical or other stress types are often a problem, and accordingly, the difficulty or ease of measurement is strongly related to the temperature value, the medium in which we desire to determine the temperature, and the overall topology of the problem [48, 60]. Some common examples of contact sensors are fiber-optic sensors [50, 51], resistors (platinum/nickel) [52, 53, 66], thermistors [67, 68], thermocouples [69, 70], cryogenic sensors [71, 72], and integrated thermometers [73, 74].

### 2.4.2. Sensors for optics

Fiber-optic sensors involve devices that are connected to various parameters using thin optical fibers as the only means of stimulating and reading the sensing element [75, 76]. These fibers are the same as those used in telecommunication devices [49, 76]. For example, when measuring the temperature in the windings of a high-voltage power transformer, the voltage can reach high values of up to 500 [kV], so the use of sensors communicating with metallic conductors is impossible for safety reasons, making this type of sensor necessary. Optical fibers have various characteristics, the variation of which can be exploited by the engineer to produce the necessary sensory instruments required for a problem [77–79]. Such characteristics are micro-bindings [80], interferometric phenomena [81, 82], changes in the refractive index [83], polarization changes [84, 85], wavelength variations [86, 87], diffractive barriers [88], and occurrence of the Sagnac effect (detection of rotational motion) [89, 90].

### 2.4.3. Sensors for electrical resistivity

The measurement of electrical resistance can lead, under the right conditions, to a fairly accurate calculation and determination of temperature. It should be pointed out here that, according to the literature, resistors and thermometers can be prepared from a wide range of materials, but the required function between electrical resistivity and temperature is not the same for all classes of materials [91, 92]. Thus, for the measurement of temperature, nickel, platinum, and copper are mostly used.

Platinum resistance thermometers (PRT) are widely used as contact sensors as most of their variants can be used for temperature measurements with an accuracy of a few [mK]. The same sensor can be used in different temperature ranges without any hysteresis effects. Its characteristics remain very stable even after many cycles of use and are characterized by low cost and high accuracy. For their activation and operation, it is necessary to have an external excitation source, which can be either current or voltage, to find the required quantity by determining their electrical resistance after a predetermined calibration procedure [93, 94].

In modern times, thin-film sensors have been established; these are electronic devices from which the wire sensors are composed of a helical very thin platinum wire placed inside the interior of a ceramic tube. In this way, protection and support of the device are achieved, and the overall cost of construction and maintenance of a system is reduced [95, 96]. In particular, wire sensors are mostly costly compared to thin-film sensors due to the purity of the metal.

To ensure the correct operation of the aforementioned devices and to avoid wear due to high thermal stresses and other environmental factors that contribute both to the destruction of the equipment and to a reduction in the accuracy of the measurement, the three-wire technique is often used [97, 98]. In particular, the operating principle is the following: we suppose three conductors, of which conductors A and B are of identical length and their resistances are at opposite ends of the bridge (cross-connection).

### 2.4.4. Thermistor sensors

One of the breakthroughs in terms of smart sensors has been thermistors. More specifically, they are made of semiconducting materials, usually metal oxides [54]. The specific conductivity of a semiconductor is given by the relation:

$$\sigma = e * (n * pe + p * ph)$$

where e is the charge of the electron; n and p are the concentrations of electron and hole carriers, respectively; and pe and ph are the electron and hole mobilities, respectively.

At this point, it is emphasized that the temperature coefficient of thermistors is generally negative, and despite the existence of thermistors with a positive temperature coefficient, its use cases are not widespread [99]. The variation in the temperature coefficient is large, which may even reach an order of magnitude of 1 percent per °C. This fact allows them to detect very small temperature changes that could not be detected by a platinum resistor or thermocouple.

Based on thermistors and the need for further analysis of the data they generate, integrated temperature sensors on semiconductors such as microprocessors were created [55, 100]. Their characteristics are the linearity of the output signal and their small size, low cost, extremely high order of accuracy, and limited operating range (from −40°C to +120°C) as long as they are satisfactorily calibrated.

Smart sensors are usually defined as remote sensors that produce their readings without being in physical contact with the system, usually by detecting the thermal radiation emitted by all available bodies with a temperature above absolute zero. As a result, in the majority of applications, this thermal radiation is detected in the infrared region of the electromagnetic spectrum [101, 102]. Their advantages are manifold as the temperatures recorded are very high and in many cases exceed the physical limits of the contact sensor materials. In addition, the difficult step of finding and designing the optimum location for sensor installation is omitted. Furthermore, wear and tear on the sensor is significantly reduced as it does not require the kind of stress that contact sensors are subjected to and also covers cases where wired contact would be impossible.

### 2.4.5. Sensors for measuring pressure

This category includes sensors used to measure the force exerted on a surface, which has the direct consequence that its unit of measurement is N·m² [56]. The pressure to be measured may be the product of liquids or gases and consists of an energy detection mechanism (Newton) and their conversion into electrical signals. The main types of these sensors are as follows: elastic pressure sensors [57], piezoelectric pressure sensors [103], and capacitive pressure sensors [104].

### 2.4.6. Rubber pressure sensors

As their name indicates, this category includes sensors whose one or more parts can be subjected to temporary changes (deformation and bending) in their dimensions [105]. These sensors are usually found in Bourdon tube pressure measurements where the operation is based on a calibrated needle placed on a surface [106]. In the event of pressure, it moves and the tube to which it is connected deviates from its initial point, and this force is measured. Due to the displacement of the needle, the above procedure is often used for distance measurement by the use of displacement sensors. Displacement refers to the change in the position of the object by some distance or angle where if it schematically depicts a straight line, it is defined as linear. Similarly, if the reference point is rotation about a given axis of rotation, it is defined as angular [107].

### 2.4.7. Capacitive pressure sensors

In this category of sensors, the diaphragm is placed between two armature elements in each of which a capacitor is formed. The two existing capacitors are then connected to a bridge, which is in equilibrium for zero pressure. The occurrence of an electrical signal disturbs the equilibrium and, therefore, changes the capacitance that contributes to the calculation of the necessary elements. The main negative aspect of these devices is that they are prone to errors in the presence of oscillations or temperature extremes. The basic structure of the measurement bridge and their structure lies in their operation, which is determined by the circuitry of the capacitors and the signal to be applied to them respectively [108–110].

### 2.4.8. Level pressure sensors

Level pressure sensors are used to control a process and are commonly found in industrial applications. In particular, they are intended to determine the maximum and minimum levels in a specific and well-defined area of action for the triggering of an actuator. If no moving parts are required in the structures concerned, they can also be converted into point-level sensors, for example, to measure capacitance or for the manufacture of lasers, infrared beams, or photocells [111, 112].

### 2.4.9. Sensors for measuring humidity

Humidity is one of the most important variables in the design and study of many elements. In particular, humidity and temperature are the main factors to be taken into account to eliminate or even find and counteract the corrosion of sensors and measurements. As far as measurement is concerned, it consists of air molecules and chemical reactions that are highly variable in the respective external environment [58, 59].

### 2.4.10. Sensors for measuring speed

In several applications, especially in terms of controlling a machine or its correct operation, it is necessary to monitor data on the flow of a process. The maintenance of airflow, for example, either for proper ventilation or to prevent overheating of a generator and heating and ventilation systems in general, is based entirely on sensors for measuring the speed of air and, in some applications, of liquids. Velocity in these measurements is defined as the distance traveled per unit of time and is expressed in meters per second [61, 113, 114].

### 2.4.11. Sensors for measuring distance

In this category, there are different implementations to achieve the same measurement depending on the objective, available budget, and desired accuracy. The first one is the sonar-type sensors where the detection and return of values are conducted using a parabolic curve in space, which covers a distance proportional to the power of the sensor. This method is preferred when it is necessary to cover a large distance between the sensor and the wall [62, 63]. Due to the mode of its operation, the measurements usually generate much noise. The second is for range sensors, where the sensor is placed at a fixed point (usually pressed) based on a fixed radius, which passes through a certain space. This beam, in the majority of cases, is light amplification by forced emission of radiation (laser) or infrared rays (infrared) [115].

### 2.4.12. Force-weight sensors

The function of weight sensors is that of the so-called S-type load cell. Essentially, it is a transducer that converts a load, in this case, a force, applied (i.e., weight) into an electrical signal [64, 116]. Installing such sensors is particularly difficult, and special attention must be paid to sensitivity, accuracy, and calibration. The operation of S-type sensors is based on the principle of the Wheatstone Bridge. In particular, the principle of operation of the bridge is to apply a potential difference to one pair of ends and measure the voltage difference. In the equilibrium state of the bridge, when no load is applied, this voltage difference is approximately equal to 0 [65].

### 2.4.13. Concise outline of sensor types

The most useful and extensively used sensor types are presented in **Table 1**.

## 3. Comparison of mini-computing solutions

After considering many well-known industry options such as Onion Omega2+ [117], ASUS Tinker Board [118], and Le Potato [119], Raspberry Pis were chosen owing to their balance of storage, speed, processor capabilities, community support, and cost-effectiveness [120]. Moreover, Omega2+ devices are less expensive and can be used in several case studies but lack processing power, whereas the Tinker Board which lacked extensive community support for sensors and documentation is proposed as an interesting solution. Similarly, from the mini-computing devices studied, Le Potato, despite its superior CPU and GPU performance, also suffered from limited community support. Given that our model of study is focused on educational purposes, it is not resource-intensive. Raspberry Pis, a solution that is not overly engineered, is in most cases suggested and preferred. All mini-computers mentioned support SD and Wi-Fi, ensuring connectivity and the ability to store local measurements cost-effectively on an SD card and transmit data remotely. A comparative analysis of these devices is provided in **Table 2**. Specifically, in order to calculate the GFLOPS, we have used the following formula:

$$\text{GFLOPS} = (\text{number of cores}) \times (\text{clock speed in GHz})$$
$$\times (\text{FLOP per cycle per core}) = C \times G \times F$$

where the F parameter is the most difficult to find out; thus, theoretically, for our measurements, we consider that many ARM cores with NEON (Advanced SIMD) instructions have the following specifications:

- Four single-precision numbers in one 128-bit NEON register per cycle
- Fused Multiply–Add (FMA), which counts as two floating-point operations (one multiply and one add).

As such, for each core, we assume that it can perform up to eight Flops per cycle (four multiplications and four additions), where, for other architectures, the vector/FMA support should be lower; thus, in our calculation, we assume four Flops per cycle.

### 3.1. Computation device signal processing and operations

As such, a question that may arise is whether the computing device alone is responsible for signal processing and all other necessary operations, given the characteristics of the sensors involved. To answer this question, one must consider the extent to which signal processing is handled by the computing device, which depends on the complexity of the processing required and the type of sensors used—that is, the system architecture.

Signal processing is managed based on the type of sensor as follows:

1. **On-sensor processing (embedded systems):** In this category, processing is performed directly on the sensor itself, through a local node, or at the edge of the sensor network. This is typical of modern sensor networks, often referred to as "smart sensors," which incorporate microcontrollers or digital signal processors to handle basic preprocessing of raw data samples and signals. The primary advantage of on-sensor processing is that it reduces the overall system load by minimizing communication with external computing devices. This approach improves energy efficiency, lowers latency, and simplifies error detection. Additionally, because it distributes processing rather than relying on a single central computing unit, it reduces the risk of a single point of failure.

2. **Edge computing processing (IoT gateways):** This processing method involves small computing devices such as Raspberry Pi, Jetson Nano, or microcontrollers like

ESP32 and STM32. These devices focus on real-time processing of data from sensors, performing tasks such as noise filtering, Fast Fourier Transform (FFT), and control algorithms. In recent years, edge devices have also been used for machine learning inference and monitoring. The key benefit of this approach is that it balances the computational load between sensors and the overall system, shifting scalable operations to the cloud, while enabling real-time analytics and horizontal scaling.

3. **Cloud server-side processing:** As the name suggests, this method involves processing data on a cloud-based middleware system or a powerful mainframe computing device. It is typically chosen for large-scale data operations, such as processing optical or industrial sensor data. Cloud computing enables deep data processing through advanced machine learning, AI-driven pattern recognition, and complex modeling algorithms. However, this approach comes with potential drawbacks, including higher power requirements, latency issues, and security concerns related to data transmission and distribution across devices.

### 3.2. Concise outline of sensors, signal processing, and their respective functionality

As such, in **Table 3**, we present a detailed analysis of the types of sensors and their respective operations in terms of signal processing, algorithms, and performed operations. For each category outlined in Section 2.4, we aim to showcase the functionality—i.e., what it does in general—and use an arrow sign to briefly indicate its main purpose and focus.

**Table 3** • A detailed analysis of the types of sensors and their respective signal processing operations and functionality

| Sensor type | Signal processing algorithms/operations | Functionality |
|---|---|---|
| Measuring temperature | <ul><li>Analog-to-Digital Conversion (ADC)</li><li>Calibration Algorithms</li><li>Noise Filtering (Kalman Filter, Moving Average)</li></ul> | Conversion of analog-to-digital temperature sensor readings.<ul><li>It reduces noise and enhances accuracy.</li></ul> |
| Optics | <ul><li>Image Processing</li><li>Fourier Transform</li><li>Edge Detection</li><li>Machine Learning</li></ul> | Detection of light intensity.<ul><li>It processes optical signals and image recognition.</li></ul> |
| Electrical resistivity | <ul><li>Wheatstone Bridge</li><li>Voltage Divider</li><li>Compensation Algorithms</li></ul> | Measurement of resistance changes.<ul><li>It detects material properties, temperature, and stress.</li></ul> |
| Thermistor | <ul><li>Exponential Curve Fitting</li><li>Signal Smoothing, ADC</li></ul> | Conversion of temperature variations into resistance changes.<ul><li>It enhances accuracy.</li></ul> |
| Measuring pressure | <ul><li>Signal Amplification</li><li>Noise Reduction</li><li>Proportional-Integral-Derivative (PID) Control</li></ul> | Converts pressure into voltage.<ul><li>It ensures stability and accuracy.</li></ul> |
| Rubber pressure | <ul><li>Strain Gauge Signal Processing</li><li>Calibration Algorithms</li></ul> | Measurement of force via material deformation.<ul><li>It is used regularly in touch-sensitive applications.</li></ul> |
| Capacitive pressure | <ul><li>Capacitance Measurement</li><li>Noise Filtering</li></ul> | Detection of pressure changes based on capacitance variations.<ul><li>It is used in medical devices, touchscreens, and industrial pressure sensing.</li></ul> |
| Level pressure | <ul><li>Differential Pressure Calculation</li><li>Signal Filtering</li></ul> | Measurement of liquid or gas levels.<ul><li>It prevents erroneous readings due to fluctuations or outliers.</li></ul> |
| Measuring humidity | <ul><li>Capacitance-Based Signal Processing</li><li>Temperature Compensation</li></ul> | Determination of air moisture content.<ul><li>It is regularly used in climate control systems.</li></ul> |
| Measuring speed | <ul><li>Pulse Counting</li><li>Fast Fourier Transform (FFT)</li><li>Doppler Effect Analysis</li></ul> | Measurement of rotational speed and velocity.<ul><li>It is used in automotive speedometers, industrial motors, and aerodynamics research.</li></ul> |

| Measuring distance | <ul><li>Time-of-Flight</li><li>Echo Processing</li><li>Laser Interferometry</li></ul> | Computation of distances using ultrasonic or optical methods.<ul><li>It is regularly used for robotics, LiDAR in autonomous vehicles, and industrial automation.</li></ul> |
|---|---|---|
| Force weight | <ul><li>Wheatstone Bridge</li><li>Load Cell Signal Processing</li><li>Low-Pass Filtering</li></ul> | Measurement of applied force or weight.<ul><li>It is used in industrial and lab settings.</li></ul> |

### 3.3. Sensor infrastructure and standards with computing devices, challenges, and open issues

#### 3.3.1. Sensors infrastructure

In this section, we focus on the interface of low-power computing devices and how they are connected to a wide range of sensors, depending on their capabilities and respective properties. The following sensor categories apply:

- **Energy and power sensors**: These include current, voltage, and energy meters used to monitor power consumption in IoT and smart grid applications. Their interfaces typically use I2C, SPI, or analog outputs. A common sensor for current measurement is the *INA219*.

- **Environmental sensors**: These typically consist of temperature, humidity, air quality, and pressure sensors, mainly used in environmental monitoring applications. Their interfaces usually use I2C, SPI, or UART to communicate with computing devices. Common examples include the *DHT11* (temperature and humidity) and *BMP280* (barometer).

- **GPS and location sensors**: These typically consist of GPS modules for positioning and tracking. They usually use UART (serial communication) to interface with computing devices. A widely used GPS module is the *NEO-6M*.

- **Motion sensors**: These typically include accelerometers, gyroscopes, and magnetometers, primarily used for motion and orientation tracking. These sensors generally communicate via I2C or SPI. A common example is the *MPU6050*, which integrates both an accelerometer and a gyroscope.

- **Optical sensors**: These typically involve image or video processing for environmental light measurements. Typical examples include Raspberry Pi devices equipped with camera modules and the *TCS3200*, a commonly used color sensor.

- **Sound sensors**: These generally consist of microphones used for sound detection or noise level measurement. Audio sensors typically require additional processing power, especially for real-time analysis. A common example is the *MAX9814*, which functions as a microphone sensor.

#### 3.3.2. Standard interfaces and interoperability of sensors

Based on the above, several standards and protocols have been established to enhance the interoperability of different sensor types within a single integrated sensor network. These include the following:

- **I2C** (Inter-Integrated Circuit): A well-established and widely used communication protocol for connecting sensors and computing devices over short distances. It is typically used for sensors that measure temperature, humidity, pressure, and acceleration.

- **PI** (Serial Peripheral Interface): A high-speed communication protocol designed for connecting devices with high-data-frequency and -throughput requirements. It is commonly used in applications requiring fast communication, such as motion sensors, cameras, and power meters.

- **UART** (Universal Asynchronous Receiver–Transmitter): A serial communication protocol often used for GPS modules, audio output sensors, and other peripherals that require asynchronous data transmission.

Beyond these three broad categories, several other notable protocols are commonly found in industrial applications and academic projects:

- **BLE** (Bluetooth Low Energy): A power-efficient wireless communication protocol used in short-range applications. It is commonly found in fitness trackers, environmental monitoring devices, and general-purpose smart home sensors.

- **Zigbee** and **LoRaWAN**: Wireless standards designed for low-power, long-range communication between devices in a sensor network. Zigbee is commonly used in home automation and industrial control applications, whereas LoRaWAN is better suited for long-range, low-bandwidth communication, particularly in rural or outdoor environments.

- **MQTT** (Message Queue Telemetry Transport): A lightweight messaging protocol designed for low-bandwidth, high-latency environments. It is widely used in IoT applications to transfer data between sensors and low-power devices.

- **IEEE 802.15.4**: A well-known standard for low-rate wireless personal area networks (WPANs), widely used in wireless sensor networks. It forms the foundation for protocols such as Zigbee and Thread.

- **OPC-UA** (Open Platform Communications Unified Architecture): A standard for secure, reliable data exchange, primarily used in industrial IoT applications to ensure interoperability between devices, sensors, and overall systems.

- **CoAP** (Constrained Application Protocol): A widely used lightweight protocol designed for constrained devices and networks. It is commonly applied in IoT environments with low-power devices and sensors.

#### 3.3.3. Challenges and open issues regarding interoperability and other key factors

Unfortunately, despite these standards and others being extensively used, interoperability remains a crucial challenge due to the following factors:

- **Protocols:** Many manufacturers still use proprietary communication protocols or data formats, making it difficult to

establish a common operational interface for sensors and their respective connected devices, especially when they come from different vendors.

- **Common data models:** Different sensor types often produce output data in proprietary formats, complicating data aggregation, storage, and ultimately, analysis. A standardized data model system is needed in industries as data integration remains a significant issue across software cycles.
- **Complexity of sensor networks:** With increase in sensor networks, managing devices that support different sets of standards, protocols, and data models becomes increasingly complex and challenging.
- **Quality of service (QoS):** Inconsistencies between sensors and network devices affect the overall capabilities of each component, leading to issues with data reliability, latency, and throughput. This, in turn, hinders the performance of the sensor network.

Lastly, beyond interoperability challenges, the following additional issues also apply:

- **Data overload and bandwidth limitations**: Communication Networks for Sensors often have limitations. In particular, a sensor network may produce large amounts of data, and transmitting these data through low-power devices to a remote repository/server/cloud system can overwhelm the existing communication network. Balancing data throughput with power efficiency is a key issue.
- **Sensor heterogeneity:** Different sensor types use various communication protocols, data formats, and power requirements, making it extremely difficult to establish a universal network where they can interface and operate within the same low-power computing devices. This challenge is usually addressed by integrating sensors with bridging technologies and implementing some level of standardization across the network.
- **Real-time data processing:** Especially in industrial control applications, health monitoring, robotics, and telemedicine, real-time processing of sensor data is often required to make time-sensitive decisions. The challenge lies in the fact that low-power devices may not be able to provide rapid responses, and latency issues can lead to system failures or operational inefficiencies.
- **Security and privacy:** Sensor networks are vulnerable to security threats, including unauthorized access, data interception, and even physical attacks. Implementing a secure and universal communication protocol with authentication mechanisms and, most importantly, data encryption for low-power devices is a significant challenge.
- **Power management:** Ensuring battery life and efficient power management is a critical issue for low-power devices, which are often battery-operated or rely on energy-harvesting techniques. To maintain long battery life and continuous data streams, it is essential to develop optimal communication networks that support monitoring and efficient power usage.

## 4. Conclusions

The current century is often characterized as the "information century," but to harness the vast amounts of information available, it is essential to understand, process, and apply data effectively to relevant problems. This article began by outlining the aim of providing young scientists, researchers, and technical hobbyists with detailed information on how to use sensory devices. We analytically defined what a sensor is, its unique characteristics, and the evolution of sensors, from simple measurement devices to smart sensors. We also elaborated on various types of sensors, emphasizing their unique capabilities and features.

The technical novelty of this article lies in presenting several core components and providing a concise literature review on a vast amount of different sensory devices and mini-computers used to develop early rapid prototypes. These prototypes can serve as a method to validate the ground truth of complex and expensive computing devices and in our case to be used as a low-power and low-cost devices to serve educational purposes. We predict that the capabilities of these devices will continue to increase, while their costs remain manageable, given their performance potential.

This article offers readers the ability to define different types of sensors, and by studying **Table 1**, they can better understand the initial steps of creating a top–down approach for their intended systems. As a result, future scientists can use this article as a reference for selecting sensors and identifying the most suitable types of sensors and mini-computers for their systems. Although Raspberry Pi is often considered the go-to solution in many cases, it is evident from the data and the tables of this manuscript that other options should also be considered based on the specific applications of each project. Lastly, as for future use cases and studies, it should be really interesting to provide a more detailed comparison between these low-cost and low-power devices and other even lower-cost devices such as Chrome Books or ChromeOS flex using devices [121].

## Acknowledgments

The authors declare no financial support for the research, authorship, or publication of this article.

## Funding

The A.P.C. was funded by *Academia Engineering* journal editors as part of a special issue/call for papers.

## Author contributions

Conceptualization, G.A. and V.T.; methodology, A.A. and Z.C.; software, P.I.; validation, G.A., A.A. and Z.C.; formal analysis, G.A.; investigation, G.A. and V.T.; resources, V.T.; data curation, G.A. and A.A. writing—original draft preparation, G.A., P.I. and V.T.; writing—review and editing, V.T., G.A. and Z.C.; supervision, G.A. and V.T.; project administration, G.A. and V.T. All authors have read and agreed to the published version of the manuscript.

## Conflict of interest

The authors declare no conflict of interest.

## Data availability statement

Data supporting these findings are available within the article, or upon request.

## Institutional review board statement

Not applicable.

## Informed consent statement

Not applicable.

## Sample availability

The authors declare no physical samples were used in the study.

## Additional information

Received: 2024-11-26

Accepted: 2025-02-06

Published:

*Academia Engineering* papers should be cited as *Academia Engineering 2025*, ISSN 2994-7065. The journal's official abbreviation is *Acad. Engg*.

## Publisher's note

Academia.edu Journals stays neutral with regard to jurisdictional claims in published maps and institutional affiliations. All claims expressed in this article are solely those of the authors and do not necessarily represent those of their affiliated organizations, or those of the publisher, the editors, and the reviewers. Any product that may be evaluated in this article, or claim that may be made by its manufacturer, is not guaranteed or endorsed by the publisher.

## Copyright

© 2025 copyright by the authors. This article is an open access article distributed under the terms and conditions of the Creative Commons Attribution (CC BY) license (https://creativecommons.org/licenses/by/4.0/).